\documentclass{aa}  
\usepackage{graphicx}
\usepackage{txfonts}
\def\kms{km\,s$^{-1}$}

\begin{document}
   \title{Multi-epoch high-resolution spectroscopy of 
   SN~2011fe\thanks{Based on observations 
  collected at the Mercator telescope, Telescopio Nazionale Galileo, 
  Nordic Optical Telescope at  Roque de los Muchachos, La Palma (Spain), and at
  the 1.82 m Copernico telescope on Mt. Ekar (Asiago, Italy).
 }}

  \subtitle{Linking the progenitor to its environment}
   \author{F. Patat\inst{1}
   \and    
    M.~A. Cordiner\inst{2} 
   \and
    N.~L.~J. Cox\inst{3} 
    \and 
    R.~I. Anderson\inst{4}
    \and
    A. Harutyunyan\inst{5}
    \and
    R.Kotak\inst{6}
    \and
    L. Palaversa\inst{4}
    \and
    V. Stanishev\inst{7}
    \and 
    L. Tomasella\inst{8}
    \and
    S. Benetti\inst{9}
    \and
    A. Goobar\inst{10}
    \and
    A. Pastorello\inst{9}
    \and
    J. Sollerman\inst{10}
}


   \institute{European Organization for Astronomical Research in the 
	Southern Hemisphere (ESO), Karl-Schwarzschild-Str.  2,
              85748, Garching b. M\"unchen, Germany
              \email{fpatat@eso.org}
              \and 
              Astrochemistry Laboratory and the Goddard Center for Astrobiology, Mailstop 691, 
              NASA Goddard Space Flight Center, 8800 Greenbelt Road, Greenbelt, MD 20770, USA
              \and
              Instituut voor Sterrenkunde, K.U. Leuven, Celestijnenlaan 200\,D, B-3001 
              Leuven, Belgium
              \and
              Observatoire de Gen\`eve, Universit\'e de Gen\`eve, 
              51 Ch. des Maillettes, CH-1290 Sauverny, Switzerland
              \and
              Fundaci\'{o}n Galileo Galilei - Telescopio Nazionale Galileo, Rambla Jos\'{e} 
              Ana Fern\'{a}ndez P\'{e}rez 7, 38712 Bre\~{n}a Baja, TF - Spain
              \and
              Astrophysics Research Center, School of Mathematics and Physics, Queens University Belfast,       	    
              Belfast, BT7 1NN, UK
              \and
              CENTRA - Centro Multidisciplinar de Astrof\'isica, Instituto 
              Superior T\'ecnico, Av. Rovisco Pais 1, 1049-001 Lisbon, Portugal
              \and
              INAF, Osservatorio Astronomico di Padova, via dell'Osservatorio 8, 
              I-36012, Asiago (VI), Italy
              \and
              INAF, Osservatorio Astronomico
               di Padova, v. Osservatorio n.5, I-35122, Padua, Italy
              \and
              Albanova  University Center, Department of Physics, Stockholm University
              Roslagstullsbacken 21, 106 91 Stockholm, Sweden
}

   \date{Received November, 2011; accepted...}

\abstract{}{}{}{}{} 
 
  \abstract
   {The nearby Type Ia SN~2011fe has provided an unprecedented  opportunity 
    to derive some of the properties  of the progenitor system, which is 
    one of the key open problems in the supernova (SN) field.}
   {This study attempts to establish a link between the reasonably well known 
   nature of the progenitor and its surrounding environment. This is done with 
   the aim of enabling the identification of similar systems in the vast 
   majority of the cases, when distance and epoch of discovery do not allow a 
   direct approach. }
   {To study the circumstellar environment of SN~2011fe we have obtained 
   high-resolution spectroscopy of SN~2011fe on 12 epochs, from 8 to 86 days 
   after the estimated date of explosion, targeting in particular at the time 
   evolution of \ion{Ca}{ii} and \ion{Na}{i}.}
   {Three main absorption systems are identified from \ion{Ca}{ii} and 
   \ion{Na}{i}, one associated to the Milky Way, one probably arising within 
   a high-velocity cloud, and one most likely associated to the halo of M101. 
   The Galactic and host galaxy reddening, deduced from the integrated 
   equivalent widths (EW) of the \ion{Na}{i} lines are $E_{B-V}$=0.011$\pm$0.002
   and $E_{B-V}$=0.014$\pm$0.002~mag, respectively. The host galaxy absorption 
   is dominated by a component detected at the same velocity measured from the 
   21-cm \ion{H}{i} line at the projected SN position ($\sim$180 \kms).  
   During the $\sim$3 months covered by our observations, its EW changed by 
   15.6$\pm$6.5 m\AA. This small variation is shown to be compatible with the 
   geometric effects produced by the rapid SN photosphere expansion coupled to 
   the patchy fractal structure of the ISM. The observed behavior is fully 
   consistent with ISM properties similar to those derived for 
   our own Galaxy, with evidences for structures on scales $\lesssim$100 AU.}
   {SN~2011fe appears to be surrounded by a "clean" environment. The lack of 
   blue-shifted, time-variant absorption features is fully consistent with the 
   progenitor being a binary system with a main-sequence, or even another 
   degenerate star.}


\authorrunning{F. Patat et al.}
\titlerunning{Multi-epoch, high-resolution spectroscopy of SN~2011fe}

   \maketitle
%

\section{\label{sec:intro}Introduction}

Since the introduction of the accreting White Dwarf (WD) scenario
(Wheelan \& Iben \cite{wi73}), the question mark about the nature of
the progenitor systems of Type Ia Supernovae (hereafter SN\,Ia) has been
growing bigger and bigger (see Patat \cite{patat11b} for a general
review on the subject). One of the most beaten tracks to obtain
information on the progenitor system (and more specifically on the
nature of the donor star) has been the search for imprints of
circumstellar material (CSM) lost by the system prior to the
explosion. Notwithstanding the extensive attempts undertaken to detect
such material, as of today no signs of it have been found in the form
of optical (Lundqvist \cite{lundqvist03,lundqvist05}; Mattila et
al. \cite{mattila05}), radio (Panagia et al. \cite{panagia06}, Chomiuk
et al. \cite{chomiuk11}), and UV/X-Ray emission (Immler et
al. \cite{immler06}).

\begin{table*}
\caption{\label{tab:log} Log of the observations} \centerline{
\begin{tabular}{ccccccccc}
\hline
Date & UT    & JD            & Phase         & Epoch     & Airmass & Exposure & Instrument & Hel. Correction \\
         &(start) &  (start) & (days)$^{(*)}$ &  (days)$^{(**)}$    & (start)  & (seconds)  &  & (km s$^{-1}$)\\
\hline
31-08-2011 & 20:39 & 2455805.357 &  $-$11   & 8.2  & 1.64 & 1800          & FIES   & $-$4.19\\
03-09-2011 & 20:20 & 2455808.344 &  $-$8    & 11.2 & 1.52 & 3600          & SARG   & $-$3.46\\
07-09-2011 & 20:20 & 2455812.340 &  $-$4    & 15.1 & 1.58 & 1800          & HERMES & $-$2.47\\
13-09-2011 & 19:23 & 2455818.308 &    +2    & 21.1 & 1.60 & 4200          & AES    & $-$0.94\\
13-09-2011 & 20:16 & 2455818.342 &    +2    & 21.1 & 1.72 & 1350          & HERMES & $-$0.96\\
14-09-2011 & 18:57 & 2455819.290 &    +3    & 22.1 & 1.50 & 3600          & AES    & $-$0.69\\
21-09-2011 & 18:59 & 2455826.291 &    +10   & 29.1 & 1.63 & 5400          & AES    & $+$1.09\\
22-09-2011 & 18:40 & 2455827.278 &    +11   & 30.1 & 1.56 & 9000          & AES    & $+$1.35\\
29-09-2011 & 20:47 & 2455834.363 &    +18   & 37.2 & 2.63 & 2700          & FIES   & $+$3.10\\
18-10-2011 & 17:54 & 2455853.246 &    +37   & 56.1 & 1.84 & 7200          & AES    & $+$7.68\\
15-11-2011 & 16:52 & 2455881.203 &    +65   & 84.0 & 2.80 & 4800          & AES    & $+$12.86\\
18-11-2011 & 01:28 & 2455883.561 &    +67   & 86.4 & 2.65 & 12600         & AES    & $+$13.52\\
18-11-2011 & 05:05 & 2455883.710 &    +67   & 86.5 & 2.92 & 4800          & SARG & $+$13.55\\
\hline
\multicolumn{9}{l}{(*) Phases from $V$ maximum were computed assuming a typical rise time of 19 days.}\\
\multicolumn{9}{l}{(**) Epochs are computed from the estimated explosion time (JD=2455797.196; Nugent et al. \cite{nugent11c}).}
\end{tabular}
}
\end{table*}

Independent information on the binary system can be obtained studying
the effects the explosion has on the companion star.  So far two
possibilities have been explored: the collision of the SN ejecta with
the companion star surface, and the stripping of the companion's
envelope.  While the former is expected to re-heat part of the SN
ejecta causing a luminosity excess in the very early epochs (Kasen
\cite{kasen10}), the latter is supposed to entrain some of the
companion atmospheric material, which should eventually become visible
in the form of emission lines at late epochs (see e.g. Marietta,
Burrows \& Fryxell \cite{marietta}, and references therein). So far,
no evidences of either phenomena have been found (Hayden et
al. \cite{hayden10}; Bianco et al. \cite{bianco11}; Leonard
\cite{leonard07}).

These facts constitute a considerable body of evidence in favor of
SN\,Ia progenitor systems where the companion star is either a MS star,
or even another degenerate body. However, it is only with the very
recent study of SN~2011fe that this was directly demonstrated with a
high level of confidence.  SN~2011fe (PTF11kly) was discovered by the
Palomar Transient Factory on August 24 UT in the outskirts of the
nearby ($\sim$6 Mpc), face-on, SAB(rs)cd galaxy M101 (NGC~5457), and
immediately classified as a very early Type Ia (Nugent et
al. \cite{nugent11a, nugent11b}). The SN was not detected (down to mag
20.6) on an image taken on August 23, implying the object was caught
within one day after the explosion.

Based on pre-explosion HST images, Li et al. (\cite{li11}) could place
for the first time a direct and stringent limit on the companion's
luminosity. This rules out a red-giant donor star, and hence a
symbiotic system as the progenitor for this particular event. The
observations presented by Li et al. (\cite{li11}) exclude the He-star
channel only partially, while a system with a MS-star transferring
mass via Roche lobe overflow is still consistent with the data, as is
a double-degenerate configuration. The lack of an early shock (Nugent
el al.  \cite{nugent11c}) goes along the same lines, ruling out
companion stars with extended atmospheres.

Thanks to the very early detection, Nugent et al. (\cite{nugent11c})
could put a stringent upper limit to the exploding star radius
($\leq$0.1 R$_\odot$), hence providing for the first time a direct
evidence that the progenitor is indeed a compact star (see also Brown
et al.  \cite{brown11}). Although a carbon-burning MS star cannot not
be excluded, a C-O WD is the most favored candidate. The range of
allowed objects was further narrowed down by Bloom et
al. (\cite{bloom11}), who derived a more stringent upper limit for the
progenitor radius (0.02 R$_\odot$), and concluded that the exploding
star must indeed be a WD (for independent considerations on the
progenitor's nature, see also Liu et al. \cite{liu11}). Bloom et
al. (\cite{bloom11}) also place an upper limit to the radius of the
companion (0.1 R$_\odot$), which excludes Roche-lobe overflowing
giants and MS stars. Finally, prompt radio and X-ray observations
placed the most stringent limit ever on the mass loss rate for a SN\,Ia:
dM/dt$\leq$10$^{-8}$ (w/100 km s$^{-1}$) M$_\odot$ yr$^{-1}$ (Horesh
et al. \cite{horesh11}).

All these facts give a very consistent view on the progenitor system
of SN~2011fe, which is very plausibly identified as a C-O WD accreting
material from a MS star (Li et al. \cite{li11}; Brown et
al. \cite{brown11}; Nugent et al. \cite{nugent11c}; Bloom et
al. \cite{bloom11}).  This is a major break-through in the
field. Having studied an object whose progenitor system is reasonably
well known, we can now ask ourselves a new question.  Would we
recognize similar systems in other SN\,Ia, whose distance and discovery
time do not allow the close scrutiny that was possible for SN~2011fe? The
key issue here is that one needs to consider the largest possible
number of observables, because, for instance, similar spectra and
light curves may come from different channels.

In this article we focus on the properties of the circumstellar
environment, whose study has led to promising results in the last
years (Patat et al. \cite{patat07a}; Simon et al. \cite{simon09};
Sternberg et al. \cite{assaf11}). For this purpose, and following the
previous studies, we have obtained multi-epoch, high-resolution,
optical spectroscopy of SN~2011fe covering the first three months of
its evolution.

The paper, reporting the results of this campaign, is structured as
follows. In Sect.~\ref{sec:obs} we give an account of the observations
and data reduction. Sect.~\ref{sec:abs} recaps the detection of narrow
absorption features, whereas Sect.~\ref{sec:ext} discusses the
implications on the extinction suffered by the SN. In
Sect.~\ref{sec:fitting} we illustrate the properties of the
intervening gas as derived from Voigt profile fittings, and we study
their behavior as a function of time. The modest time evolution is
then analyzed in Sect.~\ref{sec:ism}, in the context of the
small-scale structure of the inter-stellar medium.  The results of
this study in connection to the progenitor's nature are finally
discussed in Sect.~\ref{sec:disc}, which also summarizes our
conclusions.

\section{\label{sec:obs}Observations and data reduction}

We obtained high-resolution spectroscopy of SN~2011fe on 12 epochs,
ranging from 8 to 86 days after the estimated date of explosion
(JD=2455797.2; Nugent et al. \cite{nugent11c}). The observing log is
presented in Table~\ref{tab:log}, which includes exposure times, phase
from maximum light, epoch from explosion, airmass, and heliocentric
correction. Phases are only indicative, and were computed assuming a
typical SN\,Ia rise time of 19 days. The observations were carried out
with four different instrumental setups. These are summarized in
Table~\ref{tab:setup}, and detailed in the next subsections.

\subsection{\label{sec:aes}Asiago Echelle Spectrograph}

Seven spectra of SN~2011fe were obtained with the REOSC Asiago Echelle
Spectrograph (hereafter AES) attached to the 1.82 m telescope on
Mt. Ekar (Asiago, Italy).  The spectrograph is equipped with an an E2V
CCD42-40 AIMO, back illuminated detector (2048$\times$2048 pixel, 13.5
$\mu$m in size). A 2\arcsec\/ slit was used, with a fixed east-west
orientation, yielding a resolving power $R=\lambda/\Delta
\lambda$~$\sim$~18\,000. The wavelength range covered by the spectra
is 3780-7620~\AA.  The data reduction was performed with tasks within
the IRAF\footnote{IRAF is distributed by the National Optical
  Astronomy Observatories, which are operated by the Association of
  Universities for Research in Astronomy, under contract with the
  National Science Foundation.}  {\tt ECHELLE} package following the
  standard procedures. All images were first overscan corrected and
  trimmed. A normalized flat-field image was then derived from
  observations of a halogen lamp and used to flat field the science
  exposures. Next, each spectral order was traced and the spectra were
  extracted with the optimal extraction algorithm of Horne
  (\cite{horne86}), simultaneously subtracting the scattered
  light. The transformation from pixels to wavelength was derived from
  observations of a Thorium-Argon arc-lamp.

\begin{table}
\caption{\label{tab:setup} Instrumental setups}
\tabcolsep 1.2mm
\begin{tabular}{llcccc}
\hline
ID      &  Site    & Telescope & Res. & Disp.   & Range \\
        &          &           & power     & (\AA/pixel)  & (\AA) \\
\hline
AES     & Asiago   & 1.8-m     & 18\,000    & 0.126         & 3800-7500\\
FIES    & La Palma & 2.6-m     & 48\,000    & 0.036         & 3650-7350\\
HERMES  & La Palma & 1.2-m     & 82\,000    & 0.016         & 3770-9000\\
SARG    & La Palma & 3.5-m     & 66\,000    & 0.021         & 4600-6150\\
\hline
\end{tabular}
\end{table}

\subsection{\label{sec:fies}FIES}

Two spectra (Aug 31 and Sep 29, 2011) were obtained with the fiber-fed
(1.3\arcsec) echelle spectrograph
(FIES\footnote{http://www.not.iac.es/instruments/fies/}) mounted at
the Nordic Optical Telescope (NOT) on La Palma (Spain). We used the
medium-resolution fiber, that provides a resolving power
R$\sim$48\,000, and continuous spectral coverage from 3700 to 7300
\AA\ in a single exposure. Data were reduced using standard IRAF
tasks, as outlined in the previous section.

\subsection{\label{sec:hermes}HERMES}

Two spectra (Sep 7 and 13, 2011) were obtained with the High
Efficiency and Resolution Mercator Echelle Spectrograph (HERMES;
Raskin et al. \cite{raskin11}), mounted on the Mercator 1.2\,m
telescope in La Palma. HERMES is a high-resolution fiber-fed
(2.5\arcsec) echelle spectrograph offering spectral resolution of
$R\sim$80\,000, and full spectral coverage from 3770 to 9000 \AA\/ in
a single exposure.  Data were processed with an automated data
reduction pipeline that includes bias correction from the prescan
region on the CCD, flat-field correction via Halogen lamps, wavelength
calibration through Th-Ar reference spectra, cosmic clipping, and
background modelization.

\subsection{\label{sec:sarg}SARG}

Two spectra (Sep 3, and Nov 18, 2011) were obtained with the
Spettrografo Alta Risoluzione Galileo (SARG; Gratton et
al. \cite{gratton01}), mounted at the 3.5\,m Telescopio Nazionale
Galileo (TNG) in La Palma. SARG is a cross-dispersed echelle
spectrograph which provides spectral resolution ranging from R=29\,000
up to R=164,000.  For our observations we used the cross-disperser \#3
grism, a GG455 order sorting filter, and slit \#2 (0.8\arcsec $\times$
5.3\arcsec). This setup yields R$\sim$66,000 and a wavelength range
4600-6150 \AA\/.  The data reduction was performed using standard IRAF
tasks, as outlined in Sect.~\ref{sec:aes}.

\begin{figure}
\centerline{
\includegraphics[width=9cm]{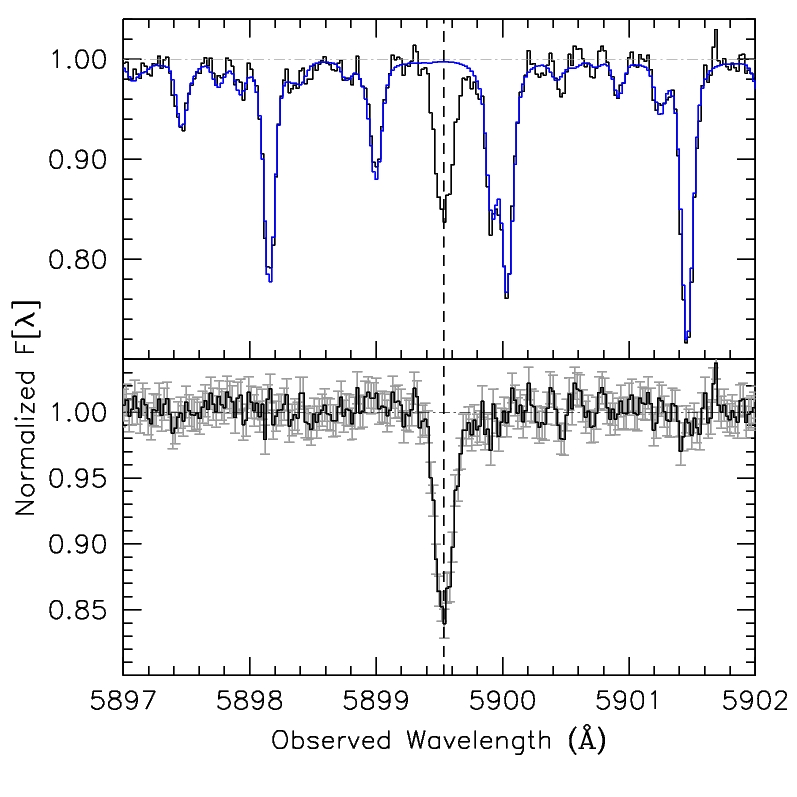}
}
\caption{\label{fig:example}Example of telluric features removal for
  the SARG spectrum in the region of \ion{Na}{i} D$_2$. Upper panel:
  observed data (black line) and best fit LBLRTM model (blue
  line). Lower panel: telluric-corrected spectrum.  The dashed
  vertical line marks the position of \ion{Na}{i} D$_2$
  ($v_{hel}\sim$180 \kms). The signal-to-noise ratio on the continuum
  is about 100 (the error-bars indicate the estimated RMS noise).}
\end{figure}

\subsection{\label{sec:telluric} Removal of telluric features}

Although the spectral region of \ion{Ca}{ii} H\&K is free of telluric
absorption features, this is not the case for \ion{Na}{i} D, and
especially \ion{K}{i} $\lambda$7665 and $\lambda$7699. Because the
observations were conducted mostly at relatively high airmasses (see
Table~\ref{tab:log}) these are quite severely disturbed by water
bands.

\begin{figure*}
\centerline{
\includegraphics[width=12.5cm, angle=-90]{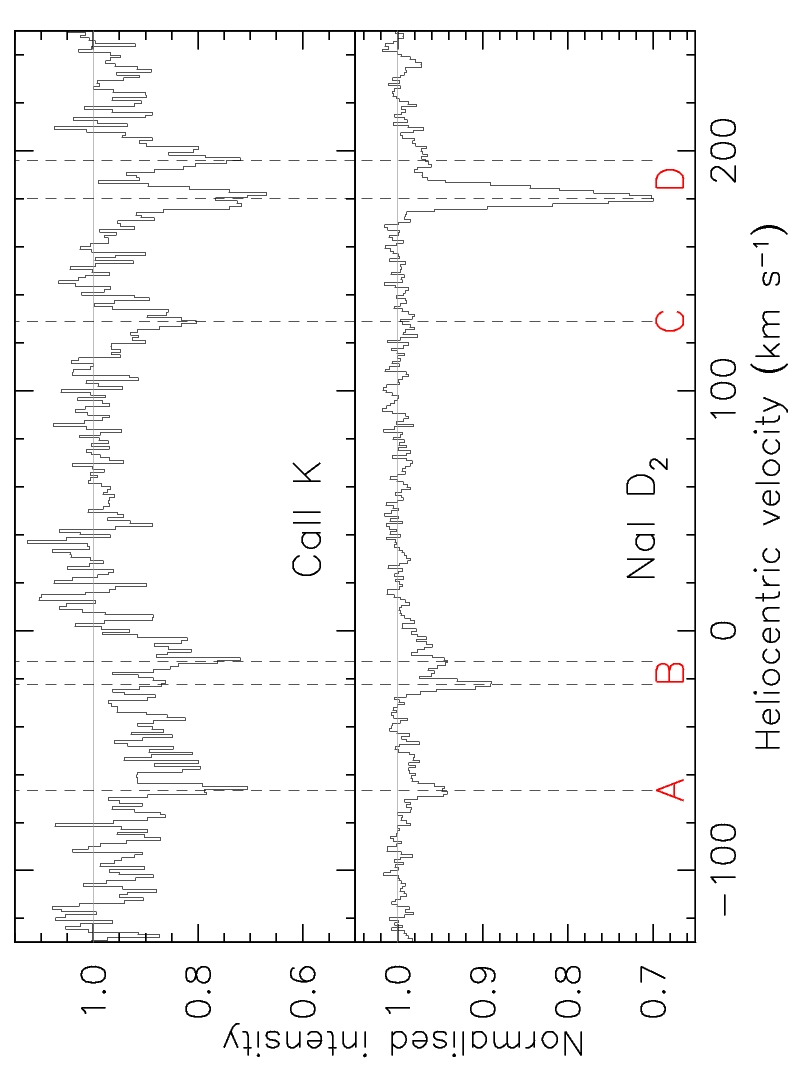}
}
\caption{\label{fig:ism}Narrow absorption components towards SN~2011fe
  for \ion{Ca}{ii} K (upper panel; day 21) and \ion{Na}{i} D$_2$
  (lower panel; day 11.2). Main absorption features are marked by
  vertical dashed lines.}
\end{figure*}

The removal of telluric absorption features is usually performed using
spectra of bright stars obtained with the same instrumental setup and
similar airmass as for the scientific target. In this paper we adopted
a different approach, using an atmospheric transmission model (see for
instance Stevenson \cite{stevenson}; Seifahrt et
al. \cite{seifahrt}). The advantages over a direct observation are: a)
the absence of intrinsic photospheric features, b) the absence of
interstellar features (chiefly \ion{Na}{i} D lines), c) the infinite
signal-to-noise, and d) the exact wavelength scale. The latter
provides an absolute reference to which the observed spectra can be
corrected via cross-correlation techniques (see below).

In this work we adopted the Line By Line Radiation Transfer Model
(LBLRTM; Clough et al. \cite{clough}). This code is based on the
HITRAN database (Rothman et al. \cite{rothman}). It has been validated
against real spectra from the UV to the sub-millimeter, and is widely
used for the retrieval of atmospheric constituents\footnote{See {\tt
    http://rtweb.aer.com/lblrtm.html}}. LBLRTM solves the radiative
transfer using an input atmospheric profile, which contains the height
profiles for temperature, pressure, and chemical composition. For our
purposes we have adopted the standard LBLRTM equatorial profile.

After computing the synthetic model in the regions of interest and for
the given airmass, this is convolved with the instrumental profile and
the wavelength correction is computed via cross-correlation. The
absence of intrinsic narrow absorptions in the SN spectrum and the
weakness of the interstellar features in the \ion{Na}{i} and
\ion{K}{i} regions (see below) make the correlation function peak very
well defined. The wavelength solutions for FIES, HERMES and SARG
spectra were all found to match the model wavelength with a maximum
deviation of 130 m s$^{-1}$, that is more than 5 times smaller than
the pixel size. Therefore, no wavelength correction was applied to
these data. In the case of the AES spectra, maximum deviations of 6.6
km s$^{-1}$ (about one pixel) were found, and the spectra were
corrected accordingly. The final wavelength accuracy is $\sim$0.1 km
s$^{-1}$ for FIES, HERMES and SARG data, while this is $\sim$0.3 km
s$^{-1}$ for the AES spectra.

An example of telluric correction is presented in
Fig.~\ref{fig:example} for the SARG spectrum. All telluric features
are removed at the level of a few percent, which is comparable to (or
better than) the continuum RMS noise of all spectra presented in this
paper.

\vspace{4mm}

As a for the sky background subtraction, we note that in some spectra
the terrestrial \ion{Na}{i} D emission components (both natural and
artificial) are not completely removed.  Because of the wavelength
shift of most features, this problem does affect only a Galactic
absorption close to null restframe velocity (see Sect.~\ref{sec:abs}),
whose profile (and thus equivalent width) is most likely affected by
the incomplete sky subtraction.

\section{\label{sec:abs} Narrow absorption features in SN~2011fe}

\subsection{\label{sec:ca} \ion{Ca}{ii} H\&K and \ion{Na}{i} D}

A number of \ion{Ca}{ii} H,K and \ion{Na}{i} D components are detected
at different velocities, as shown in Fig.~\ref{fig:ism}, which
presents the highest resolution spectra available for the two
wavelength ranges (obtained with HERMES and SARG, respectively; see
Tables~\ref{tab:log} and \ref{tab:setup}).

Two main Galactic absorption systems are detected around $-$70 and
$-$20 km\,s$^{-1}$ (marked as A and B in Fig.~\ref{fig:ism}), with
weaker and diffuse absorptions possibly present at intermediate
velocities (see the \ion{Ca}{ii} profile).  A third system is clearly
detected at about +130 km\,s$^{-1}$ (C). The recession velocity of
M101 is 241$\pm$2 km\,s$^{-1}$ (de Vaucouleurs et al. \cite{rc3}), and
therefore this component may be generated within a high-velocity cloud
(HVC) belonging either to the Milky Way or to the host galaxy. The
radial velocity derived from 21-cm line observations at the projected
SN site is about 180 km s$^{-1}$ (Bosma, Goss \& Allen \cite{bosma81};
their Fig.~1). This makes the association to M101 more plausible,
although not certain (see also Sect.~\ref{sec:fitting}).  Finally, we
detect an absorption system with two main components at about 180 and
195 km s$^{-1}$ (D). These velocities are fully consistent with the
radial velocity field at the projected SN location (Bosma et
al. \cite{bosma81}). This strongly suggests that the corresponding
absorptions arise in M101.  All features are
detected in both H and K lines, while system C is not detected in
\ion{Na}{i}. The component close to 195 km s$^{-1}$ is clearly
detected in \ion{Na}{i} only in the highest signal-to-noise spectrum
(SARG, day 11) with a total equivalent width EW=7$\pm$2~m\AA, while it
is only marginally detected in all other spectra.

Nugent et al. (\cite{nugent11c}) report the detection of one single
absorption feature (EW=45$\pm$9~m\AA), which they identify as Na~I
D$_2$ at $v$=194$\pm$1 km s$^{-1}$. Although the weighted average EW
we derive from our data (EW(\ion{Na}{i} D$_2$)=47$\pm$1~m\AA,
EW(\ion{Na}{i} D$_1$)=27$\pm$1~m\AA; see Sect.~\ref{sec:evol}) is
consistent with that reported by Nugent et al. , velocity discrepancy
is statistically very significant ($\sim$14 km s$^{-1}$), and cannot
be explained in terms of the quoted errors. Once applying the
heliocentric corrections listed in Table~\ref{tab:log} to the
different spectra, the velocities deduced for the Na~I D components
all agree to within a few 0.1 \kms. The average heliocentric velocity
deduced from the best fit profile of the most intense host galaxy
component (see Sect.~\ref{sec:fitting}) is 179.6$\pm$0.2 km
s$^{-1}$. Given the very consistent results we obtain from different
setups and epochs, we believe our result is correct to within the
reported errors.

\subsection{\label{sec:k} Other absorptions}

A careful inspection of the telluric-corrected spectra covering the
relevant wavelength range shows no trace of the two typical \ion{K}{i}
lines at 7665 and 7699 \AA, both for the Galaxy and for M101 (at the
velocities derived for the \ion{Ca}{ii} and \ion{Na}{i} features. See
previous section).  For the host galaxy the 5-$\sigma$ detection limit
derived from the HERMES spectrum taken on day 15 is EW(\ion{K}{i}
$\lambda$7699)$<$5~m\AA.  The \ion{K}{i} $\lambda$7665 is severely
affected by a telluric absorption, and so the upper limit estimate for
this line is more problematic. However, the two lines have very
similar strengths, and so the limit derived above applies to both of
them.  Similar considerations apply to the non-detection of
\ion{Ti}{ii}, \ion{Ca}{i}, CH and CH+.

Finally, none of the known strong Diffuse Interstellar Bands (DIBs;
5780, 5797, 6284, 6379, 6613; Herbig \cite{herbig95}) are detected.

\section{\label{sec:ext}Reddening}

The galactic \ion{Na}{i} D components are very weak (see
Fig.~\ref{fig:ism}), indicating that SN~2011fe suffers a very low
Galactic reddening. The integrated EW(\ion{Na}{i} D$_1$) derived from
our highest signal-to-noise spectrum (SARG, day 11) for the Galactic
components (systems A and B in Fig.~\ref{fig:ism}) is
38$\pm$5~m\AA. Applying the relation inferred by Munari \& Zwitter
(\cite{munari97}) for Galactic reddening, we get
$E_{B-V}$=0.011$\pm$0.002~mag, which is consistent with the expected
Galactic extinction along the line of sight to M101
($E_{B-V}$=0.009~mag; Schlegel, Finkbeiner \& Davies
\cite{schlegel98}). According to the Munari \& Zwitter relation, the
expected EW for the \ion{K}{i} $\lambda$7699 is $\sim$4~m\AA, which is
fully consistent with the lack of \ion{K}{i} detection in our spectra.

\begin{table*}
\tabcolsep 4mm
\centering
\caption{Na\,{\sc i} and Ca\,{\sc ii} interstellar cloud model parameters \label{tab:fits}}
\begin{tabular}{rrcccccc}
\hline\hline
\multicolumn{3}{c}{Weighted Average}&Day 8&Day 11&Day 15&Day 21&Day 37\\
\cline{1-3}
$v$&$b$&$N$&$N$&$N$&$N$&$N$&$N$\\
\hline
\multicolumn{8}{c}{\ion{Na}{i}}\\
\hline\\[-10pt]
$-67.2$&2.1&$10.52^{+0.08}_{-0.11}$&$10.52^{+0.18}_{-0.26}$&$10.53^{+0.06}_{-0.07}$&$10.36^{+0.16}_{-0.19}$&$10.62^{+0.08}_{-0.10}$&$10.50^{+0.18}_{-0.25}$\\[2pt]
$-23.3$&3.0&$10.74^{+0.07}_{-0.08}$&$10.70^{+0.13}_{-0.16}$&$10.80^{+0.04}_{-0.04}$&$10.80^{+0.07}_{-0.08}$&$10.73^{+0.06}_{-0.08}$&$10.60^{+0.16}_{-0.20}$\\[2pt]
$-12.6$&3.4&$10.55^{+0.11}_{-0.11}$&$10.18^{+0.35}_{-0.79}$&$10.64^{+0.06}_{-0.06}$&$10.27^{+0.20}_{-0.29}$&$10.66^{+0.08}_{-0.09}$&$10.80^{+0.11}_{-0.14}$\\[2pt]
$179.6$&3.8&$11.44^{+0.02}_{-0.02}$&$11.43^{+0.03}_{-0.04}$&$11.43^{+0.01}_{-0.01}$&$11.43^{+0.02}_{-0.02}$&$11.44^{+0.02}_{-0.02}$&$11.51^{+0.03}_{-0.03}$\\[2pt]
$193.6$&6.4&$10.57^{+0.12}_{-0.16}$&$10.56^{+0.20}_{-0.29}$&$10.46^{+0.09}_{-0.13}$&$10.50^{+0.16}_{-0.18}$&$10.64^{+0.11}_{-0.11}$&$10.42^{+0.26}_{-0.46}$\\[2pt]
\hline
\multicolumn{8}{c}{\ion{Ca}{ii}}\\
\hline\\[-8pt]
$-66.4$&0.7&  $11.10^{+0.17}_{-0.20}$&$11.05^{+0.34}_{-0.69}$&&$11.15^{+0.16}_{-0.21}$&$11.07^{+0.18}_{-0.24}$&\\[2pt]
$-65.7$&32.0& $11.90^{+0.07}_{-0.07}$&$12.00^{+0.08}_{-0.07}$&&$11.85^{+0.10}_{-0.10}$&$11.89^{+0.09}_{-0.10}$&\\[2pt]
$-37.5$&1.0&  $10.91^{+0.20}_{-0.24}$&$<11.10$&&$10.91^{+0.20}_{-0.30}$&$10.97^{+0.18}_{-0.25}$&\\[2pt]
$-23.5$&3.1&  $11.04^{+0.17}_{-0.24}$&$11.00^{+0.23}_{-0.35}$&&$11.03^{+0.19}_{-0.26}$&$11.03^{+0.17}_{-0.26}$&\\[2pt]
$-9.4$&7.3&   $11.63^{+0.06}_{-0.06}$&$11.72^{+0.07}_{-0.09}$&&$11.61^{+0.08}_{-0.08}$&$11.58^{+0.08}_{-0.10}$&\\[2pt]
$127.8$&6.2&  $11.54^{+0.07}_{-0.07}$&$11.61^{+0.09}_{-0.10}$&&$11.55^{+0.07}_{-0.10}$&$11.46^{+0.09}_{-0.13}$&\\[2pt]
$179.2$&5.5&  $11.82^{+0.04}_{-0.04}$&$11.89^{+0.05}_{-0.07}$&&$11.80^{+0.05}_{-0.06}$&$11.78^{+0.05}_{-0.05}$&\\[2pt]
$197.0$&6.7&  $11.58^{+0.07}_{-0.08}$&$11.64^{+0.08}_{-0.09}$&&$11.50^{+0.09}_{-0.11}$&$11.60^{+0.09}_{-0.09}$&\\[2pt]
$218.5$&3.5&  $10.89^{+0.21}_{-0.31}$&$<11.10$&&$10.99^{+0.18}_{-0.31}$&$10.82^{+0.26}_{-0.50}$&\\
\hline
\end{tabular}
\end{table*}

As shown in the previous section, we detect two features most probably
associated to the host galaxy, at $\sim$180 and $\sim$200 km s$^{-1}$,
with the latter contributing only marginally to the total \ion{Na}{i}
column density. The integrated D$_2$ EW measured on the SARG spectrum
(day 11) is 47$\pm$2~m\AA\/ (EW(\ion{Na}{i} D$_1$)=25$\pm$1~m\AA).
Applying the Munari \& Zwitter relation, this turns into
$E_{B-V}$=0.014$\pm$0.002~mag ($A_V$=0.04~mag), in agreement with the
valued derived by Nugent et al.  (\cite{nugent11c}) using the same
method.

Independent constraints on the reddening can be placed studying the
DIBs (see for instance Sollerman et al. \cite{sollerman05}, Cox \&
Patat \cite{cox08}). As it turns out, these are much less stringent
than those derived from the sodium lines, and we mention them here
only for the sake of completeness. Assuming a standard relation
between DIB strength and $E_{B-V}$ (e.g. Luna et al. \cite{luna08};
EW(6613)/$E_{B-V}$=310~m\AA), and based on the lack of DIB detection
(see previous section) gives an upper limit on the M101 reddening of
0.4~mag.  A similar assessment for the 5780 DIB (EW(5780)/$E_{B-V}$=
460~m\AA) gives E(B-V) $\leq$ 0.43~mag. Due to the intrinsic weakness
of other DIBs, the upper limits on their EW derived from our spectra
give even less stringent constraints. For example, the non-detection
of the 6379 DIB yields $E_{B-V}<$0.65~mag (EW(6379)/$E_{B-V}$=88~m\AA;
Luna et al. \cite{luna08}).

\section{\label{sec:fitting}Narrow absorption line modeling}

\begin{figure}
\centering
\includegraphics[width=8cm]{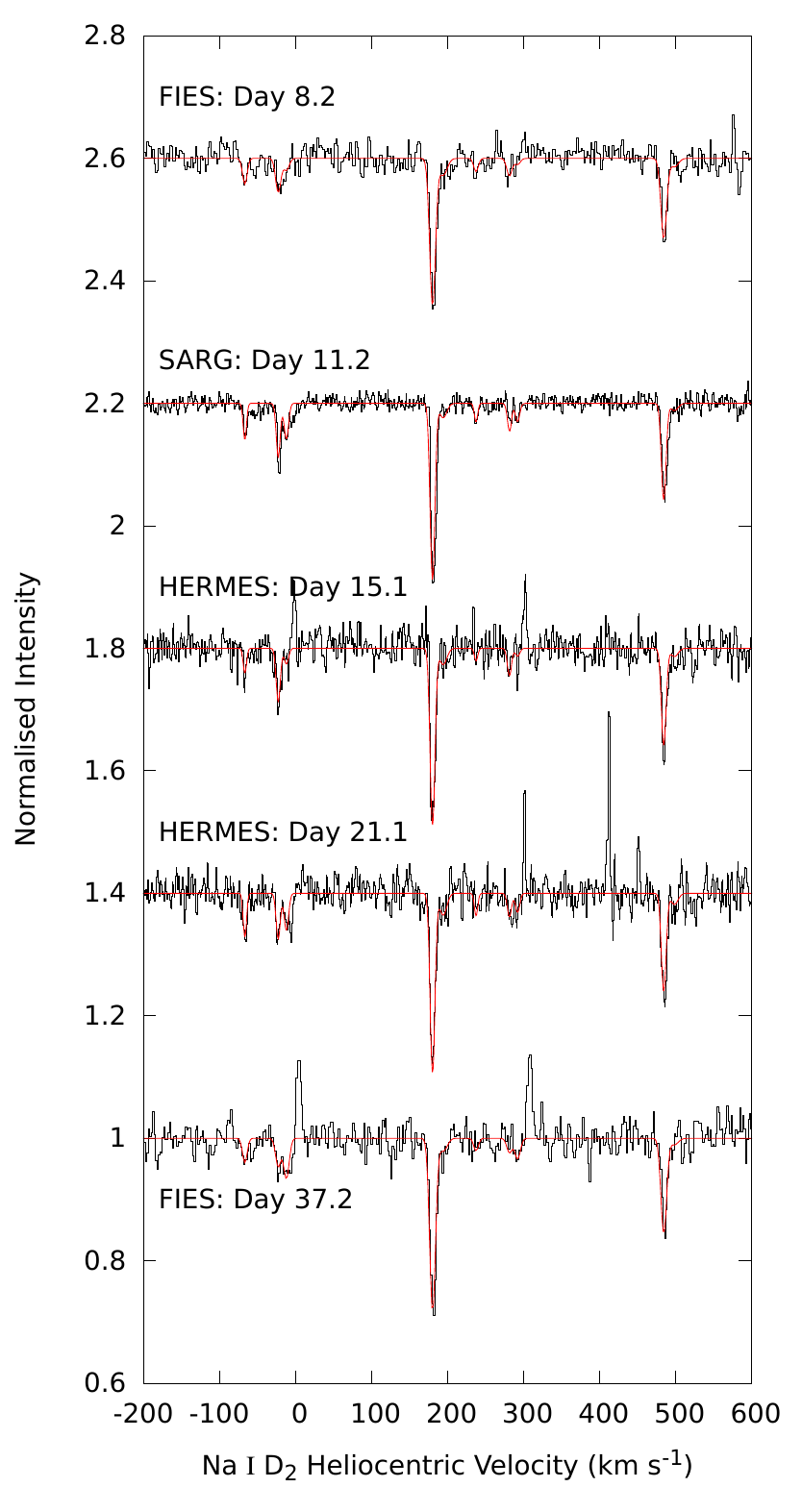}
\caption{\label{fig:naifits} \ion{Na}{i} D spectra for each epoch
  (black histograms), with best-fit models overlaid (red
  curves). Spectra have been normalised and offset vertically for
  display. Heliocentric velocity scale is given with respect to the Na
  D$_2$ line rest wavelength. The spurious emissions observed in some
  spectra around 0 and 300 \kms\/ are residuals of the sky subtraction
  (see Sect.~\ref{sec:obs}).}
\end{figure}

\begin{figure}
\centering \includegraphics[width=8cm]{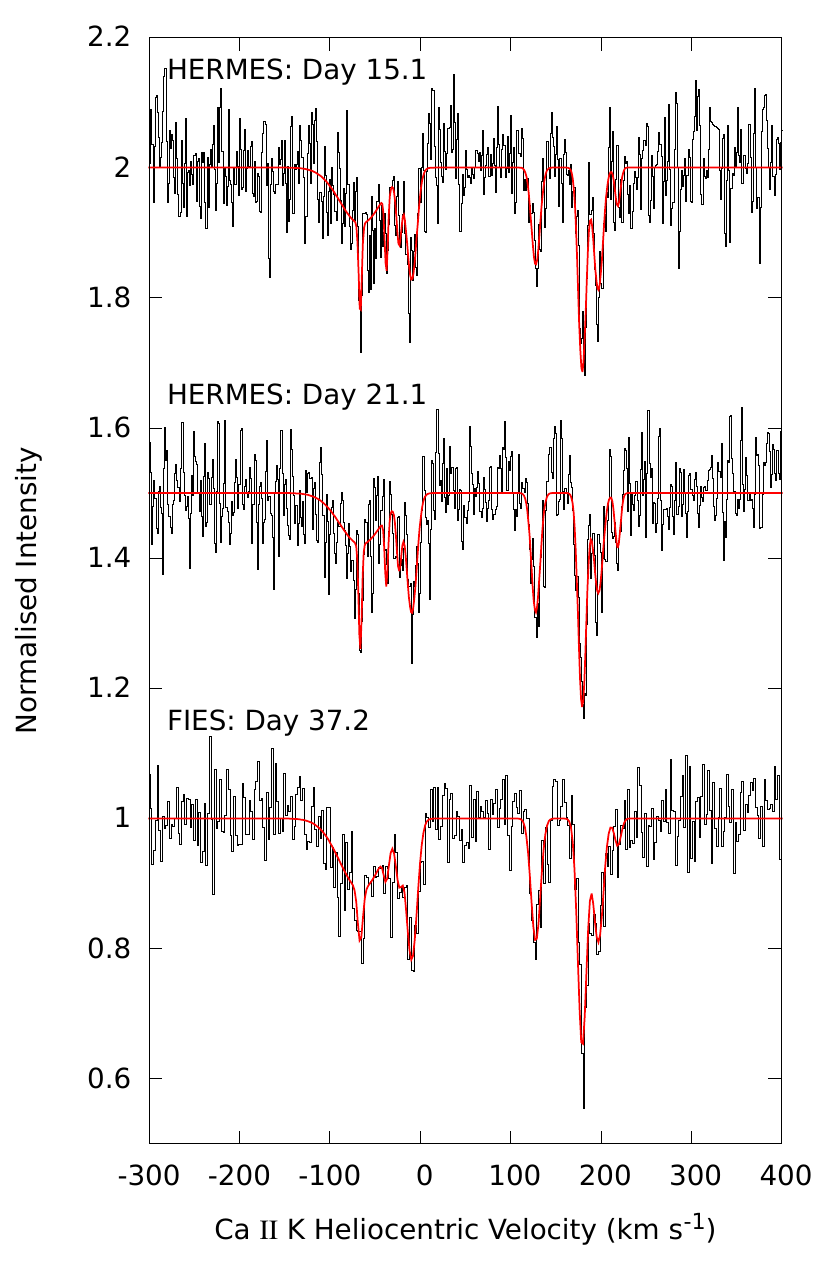}
\caption{\ion{Ca}{ii} K spectra for each epoch for which these data
  were obtained (black histograms), with best-fit models overlaid (red
  curves). Spectra have been normalised and offset vertically for
  display. Heliocentric velocity scale is given with respect to the
  \ion{Ca}{ii} K line rest wavelength.\label{fig:caiifits}}
\end{figure}

In order to derive interstellar cloud velocities, Doppler widths, and
column densities we modeled the best resolution and signal-to-noise
spectra using the {\sc vapid} routine (Howarth et
al. \cite{how02}). Saturation effects were implicitly taken into
account fitting simultaneously lines of the same species (\ion{Na}{i}
D$_1$ and D$_2$; \ion{Ca}{ii} H and K). The results for the FIES,
HERMES and SARG (first epoch only) spectra are shown in
Figs.~\ref{fig:naifits} and \ref{fig:caiifits} for \ion{Na}{i} and
\ion{Ca}{ii}. The \ion{Na}{i} spectra require 5 components for a good
fit, whereas 9 components were required for \ion{Ca}{ii}. Error
estimates were derived using Monte Carlo noise replication. The models
were re-fit to 1000 replicated data-sets, each with Gaussian noise of
RMS equal to that of the continuum added. The $\pm68$\% ranges in the
resulting data give rigorous $1\sigma$ errors on each
parameter. Results are given in Table \ref{tab:fits} for each epoch,
as well as the results of a weighted-average fit to all epochs, in
which the fitted data points were weighted by the continuum
signal-to-noise ratio of the respective spectra. The data show no
evidence for statistically significant variations in cloud component
velocities or Doppler $b$ parameters between epochs (see below), so
the $v$ and $b$ values were held fixed at the weighted-average fit
values and only the $N$ values were allowed to vary for each
epoch. Column densities (cm$^{-2}$) are given as base-ten
logarithms. The units of $v$ and $b$ are km\,s$^{-1}$.

The analysis of Milky Way interstellar gas is beyond the scope of this
paper. Here we mention that for the most marked Galactic feature
($-$23 \kms), the N(\ion{Na}{i})/N(\ion{Ca}{ii}) ratio is
$\sim$0.50, which is typical of galactic halos (Hobbs \cite{hobbs78};
Crawford \cite{crawford92}; Bertin et al. \cite{bertin93}). Sembach \&
Danks (\cite{sembach94}) measured a mean Galactic Halo
N(\ion{Na}{i})/N(\ion{Ca}{ii})$\sim$0.5.

Interestingly, the N(\ion{Na}{i})/N(\ion{Ca}{ii}) ratio for the main
feature in M101 (+180 \kms) is $\sim$0.43, indicating that it
may as well originate in the halo of the host galaxy (see also King et
al. \cite{king95} for the case of SN~1994D).  The really deviant case
is the one of the component at $\sim$128 \kms, which is clearly
detected in \ion{Ca}{ii} but not in \ion{Na}{i}.  The upper limit to
the \ion{Na}{i} column density deduced from the high-signal to noise
SARG spectrum (day 11) is 1.5$\times$10$^{10}$ cm$^{-2}$, which yields
N(\ion{Na}{i})/N(\ion{Ca}{ii})$\leq$0.04.  Such low values are rarely
observed, and indicate a mild depletion of calcium onto dust grains. A
similar case was observed for one high velocity component along the
line of sight to SN~2000cx, where \ion{Ca}{ii} was detected while
\ion{Na}{i} was not (Patat et al. \cite{patat07b}).  The corresponding
ratio was N(\ion{Na}{i})/N(\ion{Ca}{ii})$\leq$0.1.  Finally, King et
al. (\cite{king95}) found N(\ion{Na}{i})/N(\ion{Ca}{ii})=0.1--0.3 for
the high-velocity clouds observed along the line of sight to SN~1994D.
Low ratios are, in general, typical of high velocity gas (Siluk \&
Silk \cite{siluk74}). The small value we derived for the component at
$\sim$180 \kms\/ is similar to those observed in intermediate- and
high-velocity clouds by Molaro et al. (\cite{molaro93}), who concluded
that such clouds may contain very little dust, with calcium and sodium
mainly in gaseous form (see also the properties measured by Cox et
al. \cite{cox07} towards LMC sight-lines). For these reasons it is
reasonable to conclude that absorption system C (see
Fig.~\ref{fig:ism}) arises in a HVC.

\subsection{\label{sec:evol} Time evolution}

The derived cloud parameters show some marginal evidence for time
variability.  The only possibly significant variation is seen for the
host galaxy \ion{Na}{i} component at $\sim$180 km s$^{-1}$. While the
logarithmic column density derived for this absorption is
11.44$\pm$0.02 on days 8, 11, 15 and 21, on day 37 this grows to
11.52$\pm$0.03, which indicates a variation at the 3-sigma level. One
may be tempted to attribute this fluctuation to the different
instrumental setups. However, a direct comparison of the two spectra
taken on days 8 and 37 (both obtained with FIES and the same identical
setup), clearly shows that both components D$_1$ and D$_2$ had
consistently deepened by about 16\% on the second epoch, while their
FWHM remained unchanged. The integrated EWs directly measured on the
two spectra, are 23.6$\pm$1.9, 46.1$\pm$2.0~m\AA, and 30.4$\pm$2.4,
53.0$\pm$2.5~m\AA\/ for D$_1$ and D$_2$ on the two epochs
respectively, implying a variation $\Delta$EW(\ion{Na}{i}
D$_2$)=6.9$\pm$3.2~m\AA. This strengthens the conclusion that whatever
the reason for the change is, this cannot be attributed to a variation
of instrumental resolution\footnote{Being a fiber-fed spectrograph,
  FIES is not subject to changes of resolution due to seeing
  variations. Also, a check on the comparison lamp emission lines on
  the two epochs shows that the resolution indeed remained constant.}.

A similar check on the EWs can be done using the two SARG
observations, which were obtained $\sim$75 days apart (see
Table~\ref{tab:log}). The high airmass, and the limited visibility
window of SN~2011fe on the second epoch, did not allow us to reach a
very high signal-to-noise ratio ($\sim$16 at 5900\AA), so that the EW
estimate is rather uncertain.  However, the difference is
$\Delta$EW(\ion{Na}{i} D$_2$)=13.3$\pm$6.4~m\AA, which confirms the
variation at the $\sim$2-$\sigma$ level.

Although AES spectra have a significantly lower resolution (16.7 \kms,
see Table~\ref{tab:setup}), given the simple structure of the
component at $\sim$180 km s$^{-1}$, they can be used together with the
higher resolution data to study the variability on the whole time
range spanned by the available observations.  To increase the
signal-to-noise ratio we stacked the last two AES epochs (days 84 and
86).

\begin{figure}
\centering
\includegraphics[width=8cm]{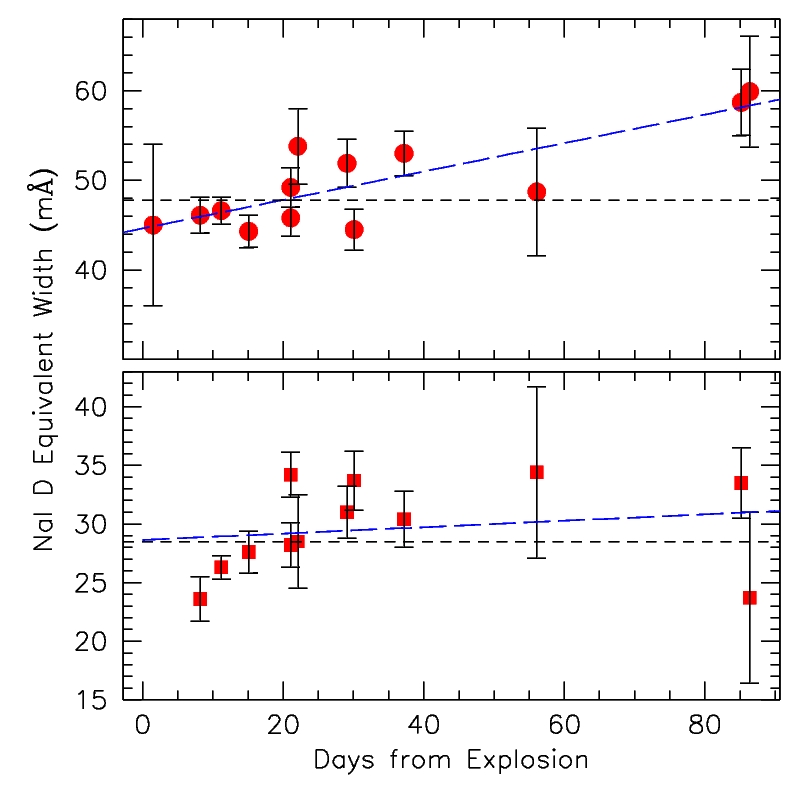}
\caption{\label{fig:ewna} EW time evolution of \ion{Na}{i} D$_2$
  (upper panel) and D$_1$ (lower panel). Values are from
  Table~\ref{tab:ew}. The short-dashed horizontal lines mark the
  weighted average values, whilst the long-dashed lines are least
  squares fits to the data.}
\end{figure}

In Table~\ref{tab:ew} we report the EWs of \ion{Na}{i} D$_2$ and D$_1$
directly measured on all our spectra, together with their associated
RMS uncertainties, the signal-to-noise ratio achieved at 5900 \AA, and
the FWHM resolution.The weighted average EWs computed over all epochs
for the two lines are 47.1$\pm$0.8~m\AA\/ and 27.4$\pm$1.1~m\AA,
respectively (the standard deviation of the EW measurements is 2.6 and
3.4~m\AA, respectively). The EWs are plotted as a function of time in
Fig.~\ref{fig:ewna}. Although all measurements are within less than
3-$\sigma$ from the weighted average (and are therefore formally
consistent with a null variation), there is a systematic increase in
both components. While the statistical significance for the D$_1$
component is low, the Pearson correlation factor is 0.82 for D$_2$.  A
least squares fit to the data gives EW(\ion{Na}{i} D$_2$)=44.6$\pm$1.4
+ (0.159$\pm$0.033) $t$, where $t$ is the time from explosion
expressed in days. The corresponding slope for D$_1$ is
0.027$\pm$0.045~m\AA\/ day$^{-1}$, which is statistically compatible
with a null variation. However, we notice that in conditions of mild
saturation (as is the case here),
EW(D$_2$)/EW(D$_1$)$\approx$1.9. Therefore, assuming the measured
slope for the D$_2$ component, the one expected for D$_1$ is
$\approx$0.08 m\AA\/ day$^{-1}$, which is not incompatible with the
low measured value.

Considering the two extreme values (days 15.1 and 86.4), the observed
peak-to-peak \ion{Na}{i} D$_2$ EW variation amounts to
15.6$\pm$6.5~m\AA\/ (considering days 15.1 and 85.2 this is
14.4$\pm$4.1~m\AA, exceeding the 3-$\sigma$ level). Although
statistically significant, this is very small when compared to the
variations seen in SN~2006X (Patat et al. \cite{patat07a}), and
SN~2007le (Simon et al. \cite{simon09}) during the first 3 months of
their evolution, which exceeded 100 m\AA\footnote{Even larger
  variations were reported for SN~1999cl (Blondin et
  al. \cite{blondin09}) and SN~2006dd (Stritzinger et
  al. \cite{strizzo10}), but no high resolution spectroscopy was
  available in those cases.}.

\begin{table}
\caption{\label{tab:ew} \ion{Na}{i} D equivalent widths}
\centerline{
\tabcolsep 1.2mm
\begin{tabular}{lccccc}
\hline
Setup & Epoch  & EW(D$_2$) & EW(D$_1$) & S/N  & Resolution \\
      & (days) & (m\AA)    & (m\AA)   & (5900\AA) & (\kms)\\ 
\hline
(*)               & 1.5  &  45.0$\pm$9.0  & -           &  -   & 6.0 \\ 
FIES\#1           & 8.2  &  46.1$\pm$2.0  & 23.6$\pm$1.9&  60  & 6.3 \\ 
SARG\#1           &11.2  &  46.6$\pm$1.5  & 26.3$\pm$1.0&  115 & 4.5 \\
HERMES\#1         &15.1  &  44.3$\pm$1.8  & 27.6$\pm$1.8&  50  & 3.7 \\
AES\#1            &21.1  &  49.2$\pm$2.2  & 34.2$\pm$1.9&  130 & 16.7\\
HERMES\#2         &21.1  &  44.3$\pm$1.8  & 27.6$\pm$1.8&  50  & 3.7\\
AES\#2            &22.1  &  53.8$\pm$4.2  & 28.5$\pm$4.0&  105 & 16.7 \\
AES\#3            &29.1  &  51.9$\pm$2.7  & 31.0$\pm$2.2&  190 & 16.7 \\
AES\#4            &30.1  &  44.5$\pm$2.3  & 33.7$\pm$2.5&  155 & 16.7 \\
FIES\#2           &37.2  &  53.0$\pm$2.5  & 30.4$\pm$2.4&  55  & 6.3 \\
AES\#5            &56.1  &  48.7$\pm$7.1  & 34.4$\pm$7.3&  50  & 16.7 \\
AES\#6+\#7        &85.2  &  58.7$\pm$3.7  & 33.5$\pm$3.0&  115 & 16.7 \\
SARG\#2           &86.4  &  59.9$\pm$6.2  & 23.7$\pm$7.3&  16  & 4.5 \\
\hline
\multicolumn{2}{l}{Weighted average}  & 47.1$\pm$0.8 & 27.4$\pm$1.1 &  &  \\
\hline
\multicolumn{6}{l}{(*)The first epoch is from Nugent et al. (\cite{nugent11c}).}\\
\end{tabular}
}
\end{table}

\section{\label{sec:ism} Implications on ISM small-scale structure}

Given the high velocity of the SN ejecta ($>$10$^{4}$ \kms), the
photosphere of a Type Ia SN expands very rapidly, reaching a radius
larger than 100 AU ($\sim$10$^{15}$ cm) at maximum light, i.e. about
20 days after the explosion. Because of this, the light an observer
receives from the SN samples different regions of intervening
inter-stellar material (ISM) as time goes by. If the ISM is patchy on
the scales of a few tens of AUs, and the density fluctuations are
large enough, this can translate into variable absorption
features. This geometrical effect and its application to the study of
ISM small scale structure are presented in Patat et
al. (\cite{patat10}), to which we refer the reader for a detailed
discussion. If we include the EW measurement published by Nugent et
al. (\cite{nugent11c}), the available data allow us to apply this
method for the first three months of the SN evolution. During this
time interval the photospheric radius increased from $\sim$10 AU to
$\sim$350 AU (the photospheric radius at our first epoch is $\sim$60
AU. See Patat et al. \cite{patat10}).

From the fact that the \ion{Na}{i} changed by $\lesssim$15~m\AA\/ over
three months, we conclude that the intervening material generating the
feature at $\sim$180 \kms\/ cannot be confined in an isolated, small
(r$<$200 AU) cloud. As the simulations show (Patat et
al. \cite{patat10}, their Fig.~3), such small clouds with central
column densities of a few 10$^{11}$ cm$^{-2}$ would be easily detected
with the time coverage and the EW accuracy characterizing our data
set. If we adopt an upper limit of 15~m\AA\/ for the variation of
EW(D$_2$), we conclude that the cloud responsible for the absorption
at $\sim$180 \kms\/ must have a radius larger than 250 AU. This does
not come as a surprise. The existence of small isolated clouds
(r$\sim$100 AU, N(\ion{Na}{i})$\sim$10$^{11}$-10$^{12}$ cm$^{-2}$), in
the past largely invoked to explain the observed \ion{H}{i} column
density fluctuations, is severely questioned in terms of pressure
equilibrium arguments, and the yet unknown mechanisms that would
produce them (see Heiles \cite{heiles97} and references therein). It
is therefore more physically meaningful to consider a diffuse ISM,
with a fractal structure described by a single power law. As it has
been shown by Deshpande (\cite{deshpande00}), this scenario provides a
convincing match to the observations down to AU scales.

The high-resolution coverage of SN~2011fe offers a unique opportunity
to derive the small scale structure properties of extra-galactic ISM,
a task which can be hardly addressed with any other known
technique. For this purpose we have computed a series of models along
the lines described in Patat et al. (\cite{patat10}), simulating the
observations of our data set, to which we have added the very early
epoch by Nugent et al. (\cite{nugent11c}). The average column density
was set to N(\ion{Na}{i})=2.8$\times$10$^{11}$ cm$^{-2}$
(corresponding to the weighted average EW(\ion{Na}{i} D$_2$)), and the
velocity parameter $b$=3.8 \kms, as deduced from the line profile
fittings. For the power-law spectral index we have used the canonical
value $\gamma$=$-$2.8 (Deshpande \cite{deshpande00}).  Given the
stochastic nature of the fractal clouds, we have adopted a statistical
approach, in which the time dependence of EW(\ion{Na}{i} D$_2$) is
examined for a large number (5000) of cloud realizations, as a
function of the peak-to-peak variation $\Delta N$ on the scale of 1000
AU.  The simulations show that peak-to-peak variations $\Delta
EW\lesssim$15~m\AA\/ imply $\Delta N\lesssim$4$\times$10$^{11}$
cm$^{-2}$ ($\Delta N/N\lesssim$1.4). With this density contrast,
maximum epoch-to-epoch variations reach 9~m\AA\/ between the first
(day 1.5) and the second epoch (day 8), with an average absolute
variation of $\sim$2~m\AA. The maximum epoch-to-epoch variations
decrease below $\sim$3~m\AA\/ already as of day 11, as the smoothing
effect of the extended SN photo-disk grows with time.  The maximum
deviation measured in our simulations between any pair of epochs in
the simulated data-set is $\sim$16 m\AA, while the 99-th percentile of
the distribution is 12 m\AA. The corresponding average absolute
variation is $\sim$4~m\AA, with a semi-interquartile range of 2
m\AA. The RMS EW variation from the weighted average is 2.6~m\AA, that
matches the value derived from the data (2.7~m\AA, see previous
section).  The simulated EW(t) curves were analyzed in detail, with
the aim of understanding whether the observed time dependency is
consistent with the fractal ISM model. They show a smooth behavior,
with the EW steadily increasing or decreasing, depending on whether
the SN appears projected onto an under- or an over-density
region. Slopes $|$dEW/dt$|\geq$0.1 m\AA\/ day$^{-1}$ are observed in
$\sim$11\% of the cases for $\Delta N/N=$1.4, while the fraction
increases to $\sim$24\% for $\Delta N/N=$2. Therefore, the relatively
large slope derived from the observations is somewhat disfavored with
respect to milder time dependencies, but still consistent with the
adopted model for the ISM structure.  For comparison, SN~2007le showed
a number of \ion{Na}{i} components that remained constant to within a
few m\AA\/ for several months (Simon et al. \cite{simon09}), implying
a null variation (i.e. $|$dEW/dt$|\lesssim$0.03 m\AA\/
day$^{-1}$). Our simulations show that for a column density
N(\ion{Na}{i})=3$\times$10$^{11}$ cm$^{-2}$ and $\Delta N /N$=1.4 this
occurs in $\sim$39\% of the realizations. For larger fluctuations
($\Delta N /N$=2) this probability decreases to $\sim$28\%. Thus, the
two different behaviors observed in SNe 2011fe and 2007le are not in
contradiction, although the latter is statistically more favored than
the former.

The derived density contrast is compatible with the values measured in
the Milky Way (Frail et al. \cite{frail94}) and, therefore, the upper
limit on the column density variation is fully in line with what is
expected for a diffuse ISM having the same properties as in our own
Galaxy. In other words, even in the case the observed fluctuations
were real (including the apparent steady growth), these are consistent
with the geometrical effect described in Patat et
al. (\cite{patat10}).  Therefore, we conclude that the absorption
feature observed at $\sim$180 \kms\/ is most likely associated to the
ISM, and is placed at distances larger than $\sim$10$^{19}$ cm
($\sim$10 light years) from the explosion site, where the effects of
the SN radiation field on the \ion{Na}{i} ionization balance are
negligible (Patat et al.  \cite{patat07b}; Simon et
al. \cite{simon09}). Unfortunately, the signal-to-noise ratio at
$\sim$3900~\AA\/ in our data is not sufficient to study with the
required accuracy the evolution of the \ion{Ca}{ii} H\&K features,
which in SNe 2006X and 2007le showed a distinct behavior (Patat et
al. \cite{patat07a}, Simon et al. \cite{simon09}).

\vspace{2mm} Because of the relatively small distance, it will be
possible to follow SN~2011fe with high-resolution spectroscopy as the
object fades away. Although this will require larger telescopes, the
data may show further variations.  As the SN is now entering its
nebular phase, it is difficult to make predictions based on our model,
because this is based on the presence of a photosphere (Patat et
al. \cite{patat10}). However, as this recedes into the ejecta and the
geometry of the emitting material changes with time, it is possible
that similar effects take place.

\section{\label{sec:disc}Discussion and conclusions}

In their analysis, Nugent et al. (\cite{nugent11c}) conclude that
SN~2011fe is a slightly sub-luminous, Branch-normal explosion, very
similar to the well studied SN~1994D (Patat et
al. \cite{patat96}). However, it is by no means clear whether all
"normal" events, sharing very similar spectroscopic and photometric
properties, do also share a common progenitor channel. For this reason
it is important to tie the [reasonably well] known progenitor nature
of this particular event to other properties, which may allow us to
identify similar events in less favorable cases, i.e. at larger
distances and with no such exceptionally early discoveries. Among
these is the circumstellar environment.

In the last five years, an alternative way of probing the explosion
environment has been deployed. The method and its first application
were presented in Patat et al. (\cite{patat07a}), and further expanded
in Simon et al. (\cite{simon09}), to which we refer the reader for a
more quantitative treatment. The idea is based on the fact that, at
variance with core-collapse events, SN\,Ia are relatively weak UV
sources. As a consequence, their ability to ionize possible
circumstellar matter is rather limited in distance. Therefore,
provided that the gas is placed along the line of sight, it can be
revealed by optical absorption lines like \ion{Ca}{ii} H\&K,
\ion{Na}{i} D, and \ion{K}{i}. These are all strong lines, and hence
they enable the detection of tiny amounts of material. As opposed to
the emission lines that are expected to arise only in the case of
direct ejecta-CSM interaction (mainly H and He), absorption lines are
generated by material placed virtually at any distance along the line
of sight.

This method has now been applied to a number of Type Ia SNe, and in
two cases has led to the detection of time-variant \ion{Na}{i}
features, which were interpreted as arising in the material lost by
the progenitor system before the explosion (Patat et
al. \cite{patat07a}; Simon et al. \cite{simon09}). The velocities of
this material are consistent with those of a red giant, hence favoring
a symbiotic system. The complex velocity structure shown by SN~2006X
was explained as being the possible outcome of recurrent-novae
episodes (Patat et al. \cite{patat07a}). This suggestion is supported
by the results of a similar analysis carried out on the known
recurrent nova RS Oph, which shows strikingly similar CSM properties
(Patat et al. \cite{patat11a}).  A study conducted on a large number
of SN\,Ia has shown a statistically significant excess of blue-shifted
\ion{Na}{i} features, which were interpreted as arising in material
lost by the progenitor system (Sternberg et al.  \cite{assaf11}). This
has given support to the conclusions reached for SNe 2006X and 2007le,
which have been suggested to be connected to binary systems hosting a
mass-losing giant. If this interpretation is correct, the measured
fraction of SN\,Ia displaying blue-shifted features suggests this
channel is not a negligible contributor to the global rate of
thermonuclear events. This conclusion is apparently inconsistent with
the limits set by the lack of ejecta-wind signatures, and the failed
detection of interaction between the ejecta and the companion
atmosphere during the very early epochs. Which may simply mean that
the picture is not complete yet; see for instance the discussion in
Chomiuk et al. (\cite{chomiuk11}) about our understanding of the wind
geometry in SN\,Ia progenitors (see also Patat \cite{patat11b}).

What makes SN~2011fe special is that the progenitor system has been
identified with a significant confidence through a number of very
stringent constraints (Li et al. \cite{li11}; Brown et
al. \cite{brown11}; Nugent et al. \cite{nugent11c}; Bloom et
al. \cite{bloom11}; Liu et al. \cite{liu11}; Horesh et
al. \cite{horesh11}). These constraints clearly rule out symbiotic
systems with a red giant companion, hence excluding recurrent novae
like RS Oph. At least for this one event.

The absence of time-variant, blue-shifted features and the ``clean''
environment reported in this paper is fully in line with the
progenitor nature derived from the very early observations. The bulk
of the material transferred from the companion remains confined to
within the binary system, and there are no recurrent outbursts that
would create the complex CSM structure seen in RS Oph (Patat et
al. \cite{patat11a}), and possibly present in SN~2006X (Patat et
al. \cite{patat07a}) and in a significant fraction of other SN\,Ia
($\geq$20\%; Sternberg et al. \cite{assaf11}). The key question is
whether the argument can be reversed, and the absence of such features
in any SN\,Ia can be interpreted in terms of a similar progenitor, at
least in a statistical sense. The answer to this question is awaiting
detailed hydrodynamical modeling, which would tell us how frequently
the physical conditions for producing observable narrow features are
met, and what the viewing angle effects are (Mohamed et al., in
preparation). Before this is accomplished, we can only speculate based
on very small numbers.

\begin{figure}
\centering
\includegraphics[width=9cm]{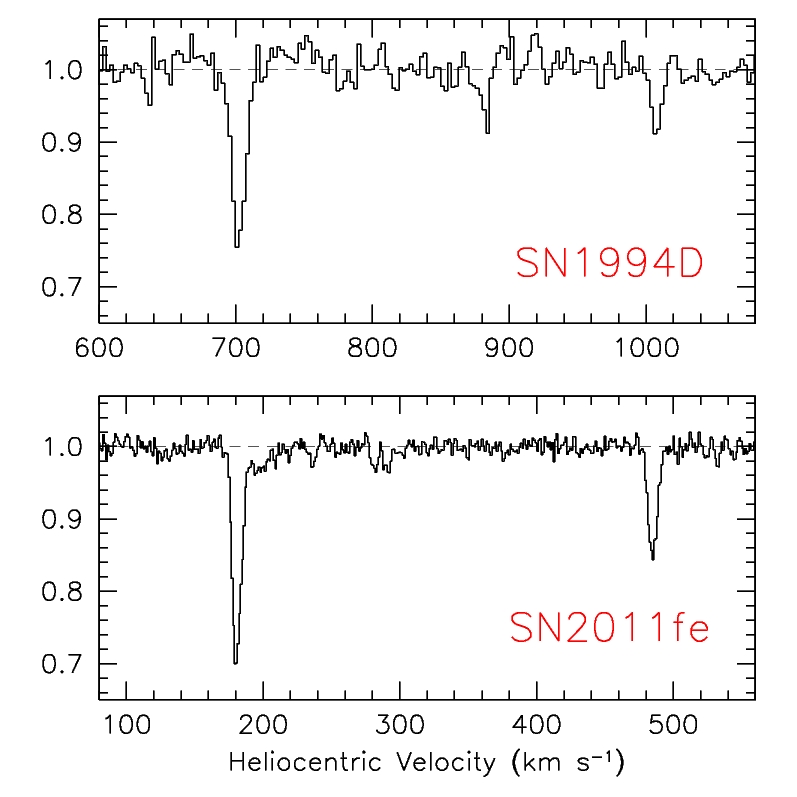}
\caption{\label{fig:sn94d}Comparison between the \ion{Na}{i} D
  spectral regions of SN~1994D (upper; Patat et al. \cite{patat96}), 
  and SN~2011fe (lower; SARG). Heliocentric velocities are computed with 
  respect to the D$_2$ component. The resolution of the SN~1994D spectrum 
  is $\sim$10 \kms.}
\end{figure}

As pointed out by Nugent et al. (\cite{nugent11c}), SN~2011fe is
spectroscopically very similar to the somewhat sub-luminous events
SN~1992A (Kirshner et al. \cite{kirshner93}) and SN~1994D (Patat et
al. \cite{patat96}). Although the link between these objects needs to
be assessed considering a number of observables, here we cannot
refrain from noticing that all three objects present a very ``clean''
circum/inter-stellar environment. To illustrate this fact, in
Fig.~\ref{fig:sn94d} we compare the high-resolution spectra of
SN~1994D and SN~2011fe in the velocity ranges corresponding to the
host galaxies recession (no measurable absorption features were
reported for SN~1992A; Kirshner et al. \cite{kirshner93}). Although
for SN~1994D there is no time coverage (see also King et
al. \cite{king95}), the two spectra are remarkably similar.

In the schema devised by Sternberg et al. (\cite{assaf11}), SNe 1994D
and 2011fe would be classified in the single/symmetric group (SS-type,
14.3\%), while SN~1992A would fall in the group with no absorption
detection (N-type, 37.1\%). An inspection of the SS group shows that 3
SS-type events (out of 5) have a single component, similar to
SN~2011fe.  If we assume that in all these cases the absorption is
attributable to the host galaxy (and not to the CSM environment), and
we add them to the N-type events, then the fraction of ``clean
events'' in the Sternberg et al. sample is $\sim$46\%. One can imagine
cases where the CSM environment is ``clean'', but the line of sight
intercepts ISM clouds that are totally unrelated to the
progenitor. Therefore, while the lack of conspicuous absorption
features can be confidently interpreted in terms of a gas-free
environment, the reverse argument does not hold.

The projected position of SN~2011fe falls close to a peripheral spiral
arm (see for instance Nugent et al. \cite{nugent11c}). However, the
very low reddening and the absence of complex absorption structures
strongly suggest the explosion took place on the front side of the
face-on host, most probably well above its galactic plane. Had the SN
been placed on the rear side of the galaxy, the IS imprints would have
been much more complex, irrespective of the progenitor's nature
deduced from the early data.  Therefore the above fraction can only
represent a lower limit. On the other hand, because of the viewing
angle effects, the fraction of events allegedly produced by systems
hosting a giant estimated by Sternberg et al. (\cite{assaf11})
is also a lower limit (20-25\%). 

The sample of SN~Ia with sufficient spectral resolution and time
coverage is still relatively small, and therefore drawing conclusions
may be premature. The fraction of events showing time-variant features
is $\sim$17\% (Patat \cite{patat11b}). Probing only the line of sight,
this indicator also suffers from viewing angle effects. In addition,
the method only reveals material placed in the suitable range of
distances from the SN, and having the required densities (Patat et
al. \cite{patat07a}; Simon et al. \cite{simon09}). Therefore, again,
it gives only a lower limit to the fraction of events generated
through this channel. Interestingly, this limit is reasonably similar
to that derived from the excess of blue-shifted features (20\%;
Sternberg et al. \cite{assaf11}), and it confirms the conclusion that
this channel contributes in a significant way to the global SN~Ia
rate.  This result has still to come to terms with the opposite
evidences given by the lack of CSM signatures in radio, UV, and X-rays
domains, and the failed detection of extended companion atmospheres in
the early stages of the SN evolution.

Although expensive in terms of telescope time, multi-epoch,
high-resolution spectroscopy of SN~Ia constitutes a powerful and
independent tool for investigating the nature of SN~Ia
progenitors. With the future availability of statistically significant
samples, it will most likely help us in settling the matter about the
contribution of the various channels leading to Type Ia
explosions. And all the pieces that currently appear to form a
contradictory set of information may fall together to form a
consistent picture.

\begin{acknowledgements}
This work has been conducted in the framework of the European
collaboration “SN Variety and Nucleosynthesis Yields”.

This work is partially based on observations made with the Mercator
Telescope, operated on the island of La Palma by the Flemish
Community, at the Spanish Observatorio del Roque de los Muchachos of
the Instituto de Astrof\'isica de Canarias.

This work is partially based on observations obtained with the HERMES
spectrograph, which is supported by the Fund for Scientific Research
of Flanders (FWO), Belgium , the Research Council of K.U.Leuven,
Belgium, the Fonds National Recherches Scientific (FNRS), Belgium, the
Royal Observatory of Belgium, the Observatoire de Geneve, Switzerland
and the Th\"uringer Landessternwarte Tautenburg, Germany.

This work is partially based on observations made with the Nordic
Optical Telescope, operated on the island of La Palma jointly by
Denmark, Finland, Iceland, Norway, and Sweden, in the Spanish
Observatorio del Roque de los Muchachos of the Instituto de
Astrofisica de Canarias.

This work is partially based on observations made with the SARG
spectrograph at the Italian Telescopio Nazionale Galileo (TNG),
operated on the island of La Palma by the Fundaci\`on Galileo Galilei of
the INAF (Istituto Nazionale di Astrofisica) at the Spanish
Observatorio del Roque de los Muchachos of the Instituto de
Astrofisica de Canarias.

This work is partially based on data collected at the 1.82 m Copernico
telescope on Mt. Ekar (Asiago, Italy).

M. C. thanks the NASA Astrobiology Institute via the Goddard Center
for Astrobiology.

V.S. acknowledges financial support from Funda\c{c}\~{a}o para a
Ci\^{e}ncia e a Tecnologia under program Ci\^{e}ncia 2008 and a
research grant PTDC/CTE-AST/112582/2009.

\end{acknowledgements}

\end{document}